\documentclass[a4paper,12pt]{article}

\usepackage{ifpdf}

\newif\ifpdf
\ifx\pdfoutput\undefined
  \pdffalse
\else
  \pdfoutput=1
  \pdftrue
\fi

\RequirePackage{xspace} %
\RequirePackage{subfigure} %
\RequirePackage[centertags]{amsmath} %
\RequirePackage{amssymb}
\RequirePackage{wrapfig} %
\RequirePackage{calc} %
\RequirePackage{ifthen}
\RequirePackage{tabularx} %
\RequirePackage{flafter} %
\RequirePackage{fancyhdr} %

\ifpdf
  \RequirePackage[pdftex]{color}%
  \RequirePackage{colortbl}%
  \RequirePackage{array}%
  \RequirePackage[pdftex]{graphicx}

  \RequirePackage[ pdftex, plainpages = false, pdfpagelabels,
                 pdfpagelayout = useoutlines,
                 bookmarks,
                 breaklinks = true,
                 linktocpage,
                 pagebackref,                      
                 colorlinks = true,
                 linkcolor = blue,
                 urlcolor  = blue,
                 citecolor = blue,
                 anchorcolor = blue,
                 hyperindex = true,
                 hyperfigures
                 ]{hyperref}

\else
  \RequirePackage{color}
  \RequirePackage{colortbl}
   \RequirePackage{array}
  \RequirePackage[dvips]{graphicx}
  \RequirePackage{hyperref}
  \usepackage{rotating}
\fi

\usepackage{makeidx} 
\usepackage{setspace} 
\usepackage{rotating} 
\usepackage{ecltree}
\usepackage{epic}
\usepackage{supertabular}  
\usepackage{color}
\usepackage{exscale}
\usepackage{fontenc}
\usepackage{ifthen}
\usepackage{latexsym}
\usepackage{makeidx}
\usepackage{syntonly}
\usepackage{inputenc}
\usepackage{graphicx}
\usepackage{setspace}
\usepackage{caption2}
\usepackage[english]{babel}
\usepackage[square, comma,numbers,sort&compress]{natbib}
\usepackage{hypernat}
\usepackage{boxedminipage}
\usepackage{framed}
\usepackage{longtable}
\usepackage[all]{hypcap}

\newcommand{\note}[1]{\marginpar[left]{\singlespace \tiny #1}}

\renewcommand{\sectionmark}[1]%
      {\markright{\thesection\ #1}} 

\renewcommand{\note}[1]{}

\doublespace 

\begin{document}

\pagenumbering{arabic}

\pagestyle{headings} %
\addtolength{\headheight}{+1.6pt}
\lhead[{Chapter \thechapter \thepage}]%
      {{\bfseries\rightmark}}
\rhead[{\bfseries\leftmark}]%
     {{\bfseries\thepage}} 
\headsep = 1.0cm               

\begin{center}
{\Large Accounting for the Use of Different Length Scale Factors in $x$, $y$ and $z$ Directions }
\par\end{center}{\Large \par}

\begin{center}
Taha Sochi (taha.sochi@kcl.ac.uk)
\par\end{center}

\begin{center}
Imaging Sciences \& Biomedical Engineering, King\textquoteright{}s College London, The Rayne
Institute, St Thomas\textquoteright{} Hospital, London, SE1 7EH, UK
\par\end{center}

\begin{center}
Abstract: This short article presents a mathematical formula required for metric corrections in
image extraction and processing when using different length scale factors in three-dimensional
space which is normally encountered in cryomicrotome image construction techniques.
\par\end{center}

\begin{center}
\par\end{center}

In many scientific and industrial situations, the coordinates space is scaled by different length
factors in the three spatial directions, $x$, $y$ and $z$; which affect the metric relations. For
instance, in the cryomicrotomic image extraction techniques, the thickness of slices may be subject
to errors or variations making the voxel size in one direction larger or smaller than its standard
size in the two other directions. Consequently, the geometric parameters obtained from these
images, which are based on the standard units of image space of an assumed cubic voxel unit, will
be contaminated with errors causing a distortion because of the missing scale factors required by
the isotropy of the physical space.

In the following we present a simple case based on real-life cryomicrotomic image construction
algorithms in biomedical applications where vasculature trees are obtained by computing the radius
of each vessel in a number of rotational steps through a whole circle and the results are then
averaged to obtain the final radius [1, 2]. As these rotational steps are oriented differently in
the 3D space, the contribution of the length scale factors will vary from one orientation to the
other and hence a scaling correction is required to obtain the correct radius.

There are several possible ways for deriving a formula for this correction; these include the use
of rotational matrices, and circle projection on the three standard planes (i.e. $xy$, $yz$ and
$zx$) to obtain three ellipses from which the three spatial components can be computed. However, an
easier and more efficient way is to find a parameterized form of an intersection circle between a
plane perpendicular to the vessel axis and the vessel itself, which in essence is equivalent to a
great circle intersection of this plane and a sphere having the same radius as the vessel [3]. This
method of derivation is outlined below.

For a regular cylindrical straight vessel oriented arbitrarily in 3D space and defined by its two
end points $P_{1}(x_{1},y_{1},z_{1})$ and $P_{2}(x_{2},y_{2},z_{2})$ and radius $r$, a free vector
oriented in its axial direction is given by

\begin{equation}
\boldsymbol{a}=(a_{x},a_{y},a_{z})=(x_{1}-x_{2},y_{1}-y_{2},z_{1}-z_{2})\end{equation}

while a plane perpendicular to this vector and (for simplicity with no loss of generality) passing
through the origin is given by

\begin{equation}
a_{x}x+a_{y}y+a_{z}z=0\end{equation}

Now, in 3D space a parameterized circle of radius $r$ centered (with no loss of generality) at the
origin and lying in a plane identified by two orthonormal vectors $\boldsymbol{b}$ and
$\boldsymbol{c}$ is given by the equation

\begin{equation}
r\left[\cos(t)\boldsymbol{b}+\sin(t)\boldsymbol{c}\right]\,\,\,\,\,\,\,\,\,\,\,\,0\le
t<2\pi\end{equation}

that is

\begin{equation}
\left(r\left[\cos(t)b_{x}+\sin(t)c_{x}\right],r\left[\cos(t)b_{y}+\sin(t)c_{y}\right],r\left[\cos(t)b_{z}+\sin(t)c_{z}\right]\right)\,\,\,\,\,\,\,\,\,\,\,\,0\le
t<2\pi\end{equation}

To find $\boldsymbol{b}$ and $\boldsymbol{c}$, a formal orthogonalization process, such as
Gram\textendash{}Schmidt, with normalization can be followed where random vectors non-collinear to
$\boldsymbol{a}$ can be used. However a more convenient way is to find an arbitrary non-trivial
vector lying in the plane by inserting arbitrary values for two variables (e.g. $x=1$ and $y=1$ )
in the plane equation and solving for the other variable ($z$) followed by normalizing through the
division by its norm. If this vector is considered $\boldsymbol{b}$, then vector $\boldsymbol{c}$
is found by taking the cross product $\boldsymbol{a}\times\boldsymbol{b}$ and normalizing.

If the following length scale factors: $\alpha$, $\beta$ and $\gamma$ are introduced on the $x$,
$y$ and $z$ directions respectively, then the distorted radius, $r'$, at a random orientation
$t=\theta$ is given by

\begin{equation}
r'=r\sqrt{\left(\alpha\left[\cos(\theta)b_{x}+\sin(\theta)c_{x}\right]\right)^{2}+\left(\beta\left[\cos(\theta)b_{y}+\sin(\theta)c_{y}\right]\right)^{2}+\left(\gamma\left[\cos(\theta)b_{z}+\sin(\theta)c_{z}\right]\right)^{2}}\end{equation}

and hence the actual radius, $r$, is given by

\begin{equation}
r=\frac{r'}{\sqrt{\left(\alpha\left[\cos(\theta)b_{x}+\sin(\theta)c_{x}\right]\right)^{2}+\left(\beta\left[\cos(\theta)b_{y}+\sin(\theta)c_{y}\right]\right)^{2}+\left(\gamma\left[\cos(\theta)b_{z}+\sin(\theta)c_{z}\right]\right)^{2}}}\end{equation}

As the image construction algorithm computes $r'$ at $N$ rotational steps (e.g. 360 steps
corresponding to $360^{\circ}$) and averages the results, to restore the corrected radius $r$, this
correction should be introduced at each one of these steps. In an ideal situation where rotational
symmetry holds, only one quarter of these steps, $\frac{N}{4}$, is required, resulting in a
substantial computational economy. However due to the measurements and algorithmic errors at each
step, it may be safer to maintain the $N$ steps as the errors are expected to level out or diminish
by applying this process through the whole circle.

This correction can also be extended to use for post processing correction by applying the
correction on the final averaged radius following a correction-free extraction process. For $N$
rotational steps we have

{\small \begin{equation}
\sum_{i}^{N}r'_{i}=r\sum_{i}^{N}\sqrt{\left(\alpha\left[\cos(\theta_{i})b_{x}+\sin(\theta_{i})c_{x}\right]\right)^{2}+\left(\beta\left[\cos(\theta_{i})b_{y}+\sin(\theta_{i})c_{y}\right]\right)^{2}+\left(\gamma\left[\cos(\theta_{i})b_{z}+\sin(\theta_{i})c_{z}\right]\right)^{2}}\end{equation}
}{\small \par}

Since the averaged post processing radius is

\begin{equation}
R_{av}=\frac{\sum_{i}^{N}r'_{i}}{N}\end{equation}

the actual radius is then given by

\begin{equation}
r=\frac{NR_{av}}{\sum_{i}^{N}\sqrt{\left(\alpha\left[\cos(\theta_{i})b_{x}+\sin(\theta_{i})c_{x}\right]\right)^{2}+\left(\beta\left[\cos(\theta_{i})b_{y}+\sin(\theta_{i})c_{y}\right]\right)^{2}+\left(\gamma\left[\cos(\theta_{i})b_{z}+\sin(\theta_{i})c_{z}\right]\right)^{2}}}\end{equation}

Although post processing correction may not result in computational efficiency, it may be more
convenient and useful to use when the non-corrected data are already obtained with no requirement
to repeat the extraction process.

It should be remarked that this correction can be applied in general to correct for this type of
distortion regardless of the number of steps (single or multiple) and the shape of the object as
long as the $x$, $y$ and $z$ components of the position vector can be obtained for each point in
space required to trace the path of the distorted shape. This process can also be extended from
discrete to continuous by substituting the summations with integrations with some other minor
modifications to account for this correction in analytical contexts rather than numerical discrete
processes.

\clearpage

\textbf{References}

[1] R.D. ter Wee, H. Schulten, M.J. Post, J.A.E. Spaan. Localization and visualization of
collateral vessels by means of an imaging cryomicrotome. Vascular Pharmacology, 45(3): e63-e64,
2006.

[2] B. Bracegirdle. A History of Microtechnique: The Evolution of the Microtome and the Development
of Tissue Preparation. Science Heritage Ltd, 2nd edition, 1986.

[3] G.B. Thomas, R.L. Finney. Calculus and Analytic Geometry, Addison Wesley, 9th Edition, 1995.

\end{document}